\documentclass[fleqn,10pt]{wlscirep}
\usepackage[utf8]{inputenc}
\usepackage[T1]{fontenc}
\usepackage{setspace}
\usepackage{bm}
\usepackage[]{lineno}
\usepackage{subcaption}

\usepackage{verbatim}

\usepackage{siunitx}
\let\DeclareUSUnit\DeclareSIUnit
\let\US\SI
\DeclareUSUnit\inch{in}

\newcommand\wordcount[1]{
    \immediate\write18{texcount -sub=section \jobname.tex  | grep "Section" | sed -e 's/+.*//' | sed -n \thesection p > 'count.txt'}
(\input{count.txt}words)}

\usepackage{silence}
\title{Level set methods for gradient-free optimization of metasurface arrays}

\definecolor{impactpurple}{cmyk}{81, 99, 0, 0}
\definecolor{canopylime}{cmyk}{38, 0, 94, 0}

\DeclareFontFamily{U}{BOONDOX-calo}{\skewchar\font=45 }
\DeclareFontShape{U}{BOONDOX-calo}{m}{n}{
  <-> s*[1.05] BOONDOX-r-calo}{}
\DeclareFontShape{U}{BOONDOX-calo}{b}{n}{
  <-> s*[1.05] BOONDOX-b-calo}{}
\DeclareMathAlphabet{\mathcalboondox}{U}{BOONDOX-calo}{m}{n}
\SetMathAlphabet{\mathcalboondox}{bold}{U}{BOONDOX-calo}{b}{n}
\DeclareMathAlphabet{\mathbcalboondox}{U}{BOONDOX-calo}{b}{n}

\newcommand{\R}{\mathbb{R}}

\newcommand{\Z}{\mathbb{Z}}

\newcommand{\vct}[1]{\bm{#1}}
\newcommand{\mtx}[1]{\bm{#1}}

\DeclareMathOperator*{\minimize}{\text{minimize}}
\DeclareMathOperator*{\maximize}{\text{maximize}}

\newcommand{\vc}{\vct{c}}

\newcommand{\vs}{\vct{s}}
\newcommand{\vt}{\vct{t}}

\newcommand{\vx}{\vct{x}}
\newcommand{\vy}{\vct{y}}
\newcommand{\vz}{\vct{z}}
\newcommand{\valpha}{\vct{\alpha}}
\newcommand{\vbeta}{\vct{\beta}}

\newcommand{\mG}{\mtx{G}}
\newcommand{\mH}{\mtx{H}}

\author[1,*,+]{Alex Saad-Falcon}
\author[1,2,+]{Christopher Howard}
\author[1]{Justin Romberg}
\author[1,2]{Kenneth Allen}

\affil[1]{Georgia Institute of Technology, School of Electrical and Computer Engineering, Atlanta, GA 30332 USA}
\affil[2]{Georgia Tech Research Institute, Advanced Concepts Laboratory, Atlanta, GA 30318 USA}

\affil[*]{Corresponding author: alexsaadfalcon@gatech.edu}
\affil[+]{these authors contributed equally to this work}

\begin{abstract}
Global optimization techniques are increasingly preferred over human-driven methods in the design of electromagnetic structures such as metasurfaces, and careful construction and parameterization of the physical structure is critical in ensuring computational efficiency and convergence of the optimization algorithm to a globally optimal solution. While many design variables in physical systems take discrete values, optimization algorithms often benefit from a continuous design space. This work demonstrates the use of level set functions as a continuous basis for designing material distributions for metasurface arrays and introduces an improved parameterization which is termed the periodic level set function. We explore the use of alternate norms in the definition of the level set function and define a new pseudo-inverse technique for upsampling basis coefficients with these norms. The level set method is compared to the fragmented parameterization and shows improved electromagnetic responses for two dissimilar cost functions: a narrowband objective and a broadband objective. Finally, we manufacture an optimized level set metasurface and measure its scattering parameters to demonstrate real-world performance.

\end{abstract}

\begin{document}

\newgeometry{top=1.5in, bottom=1in, left=1in, right=1in}
\thispagestyle{empty}
\restoregeometry

\flushbottom
\maketitle
\thispagestyle{empty}
\setcounter{page}{1}


\section*{Introduction}

Just as the intrinsic properties of any material are determined by the chemical and physical arrangement of constitutent atoms, the metamaterial is an artifical material composed of an arrangement of "meta-atoms"\cite{dudley2017} -- sub-wavelength engineered structures tailored to impart an amplitude and phase change upon an incident electromagnetic wave. While any number of distinct meta-atoms might be flexibly arranged to create a wide range of spatial and frequency responses as in disordered\cite{vynck2012} or multifunctional\cite{kim2021} metamaterials, the resulting complexity due to field coupling between meta-atoms typically demands that a single meta-atom or unit cell is designed and infinitely replicated across a periodic array to form a homogenized metamaterial.\cite{so2023} The two-dimensional subset of bulk metamaterials are metasurfaces, where tailoring surface impedance of a patterned metal/dielectric layer customizes the reflection and transmission of an incident wave in amplitude, phase, and polarization.\cite{chen_review_2016} Metamaterials and metasurfaces have been shown to demonstrate unique physical properties, including left-handedness,\cite{shelby2001} electromagnetic band-gap behavior,\cite{rahmatsamii2001ebg} frequency selectivity,\cite{ali_metamaterials_2022} nonlinearity\cite{almeida2016}, and polarizability.\cite{shi2015dual} Many years of intuition-based design have yielded a library of canonical metasurfaces, e.g., the split-ring resonator,\cite{islam2022} square loop,\cite{kent2010new} Jerusalem cross,\cite{monavar2011bandwidth} etc., capable of producing these properties. However, there is no expectation that these designs -- a small subset of the entire possible design space of materials arranged in the electromagnetic volume of the meta-atom -- represent the optimal solution. Although fundamental limits on transmission and reflection varying over angle, frequency, and polarization can be derived from causality and other physical first principles,\cite{arbabi2017fundamental,molesky2018inverse} no standard design process exists to direct a metasurface's performance to these fundamental limits.



Design of increasingly advanced metasurfaces therefore benefits from global topology optimization techniques to explore the design space for an optimal solution. Design space parameterization is required because the computational cost of simulations limits the possible number of cost function evaluations, which in turn significantly limits the number of design variables which can be tuned by optimization algorithms. For example, the canonical metasurface shapes described above have few design variables which are optimized quickly, but only locally optimal solutions are attained.

\begin{figure}[ht] 
\begin{center}
\includegraphics[width=\linewidth]{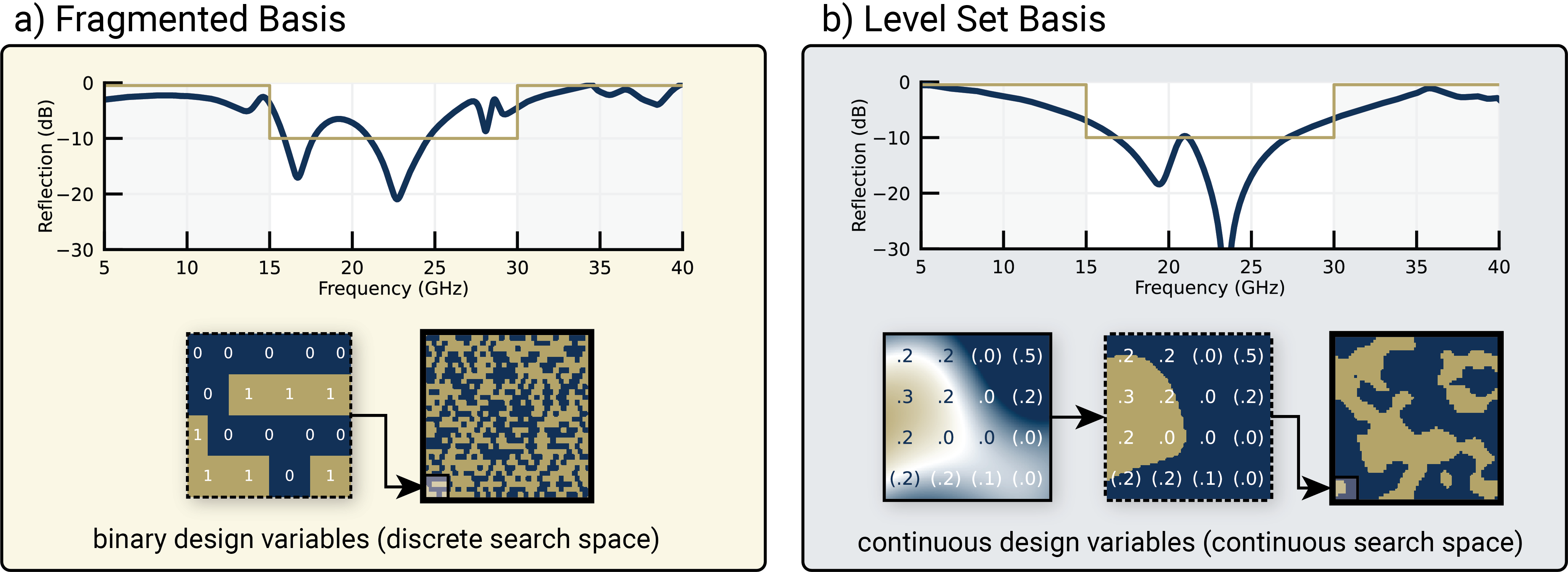}
\caption{Metasurface optimization is performed with fragmented and periodic level set function (P-LSF) parameterizations for topology optimization of a metasurface. The objective for both designs is a bandpass filter from 15-30~GHz.
(\textbf{a}) The fragmented parameterization consists of binary-valued design variables which determine the presence or absence of metal at each grid location.
(\textbf{b}) The P-LSF parameterization consists of real-valued design variables, and metal is placed at locations where the P-LSF values are greater than 0. The P-LSF design achieves improved bandpass performance.
}
\label{fig:overview}
\end{center}
\end{figure}

A more expressive design space parameterization for metamaterials is the ``fragmented'' or ``coded'' basis\cite{friederich2001, cui_coding_2014} shown in \autoref{fig:overview}(a). The material distribution function for fragmented structures is parameterized by a set of binary design variables, with each variable determining the presence or absence of material on a grid. While the fragmented parameterization offers the designer clarity in its one-to-one mapping of parameters to material features, discrete design variables used for material placement are necessarily non-convex and hence detrimental to many optimizers.\cite{rockafellar_lagrange_1993, boyd_convex_2004}

Evolutionary approaches such as genetic algorithms\cite{allen2020, zhang_genetic_2021} (GAs) and particle swarm optimization\cite{rahmat-samii_nature-inspired_2012} (PSO) have been used to optimize fragmented structures. Other nature-inspired optimizations demonstrated on metamaterials include ant colony optimization\cite{zhu2017design} (ACO) and wind driven optimization\cite{bayraktar2010wind} (WDO).
Given the sample inefficiency of these optimizers, it is sometimes desirable to use more judicious strategies such as Bayesian optimization. One approach proposed by Hou et al.\cite{hou_compression_2021} is to first remap the binary design variables to a set of continuous design variables using a compression mapping and then apply Bayesian optimization to the new variables. However, the inference time in Bayesian optimization grows $O(N^3)$ with the number of samples $N$, which restricts the total number of function evaluations to a few hundred ($N=200$ in Hou et al. \cite{hou_compression_2021}). Bayesian optimization is also a sequential design strategy, so multiple function evaluations cannot be run in parallel to take advantage of high-performance computing developments.

The convergence of metamaterial optimizations may be accelerated by employing adjoint methods to compute the gradient of the cost function with respect to the design variables. Used in conjunction with gradient-based optimizers, adjoint methods quickly converge to an optimum.\cite{elsharabasy_wide-angle_2020, hammond_high-performance_2022, schubert_inverse_2022} However, adjoint analysis has not been widely adopted in commercial electromagnetics solvers, and in some cases the adjoint operator is simply not computable, highlighting a need for robust gradient-free optimization techniques. Surrogate-based optimization \cite{zhang_genetic_2021, pestourie_active_2020} (SBO) precludes the need to derive an adjoint operator, and a cheap forward or inverse model is learned from data instead. Surrogate models have challenges with model architecture, data representation, and generalization. But more prohibitively, tens of thousands of samples, each generated by a computationally expensive full-wave electromagnetic simulation, are required to train an accurate model for high-dimensional nonlinear cost functions.\cite{alizadeh_managing_2020}

Level set functions (LSFs) are a promising continuous basis for gradient-free optimization of metamaterials. LSFs have been applied to gradient-based optimization for electrostatic systems,\cite{young_sun_kim_level_2009} antennas,\cite{zhou_level-set_2010}, acoustic metamaterials,\cite{noguchi2021,guo2023} nanophotonic devices,\cite{vercruysse2019} and multi-port microwave networks.\cite{murai_multiscale_2023} Mansouree and Arbabi\cite{mansouree2019} leverage the level-set method and gradient descent techniques to design an optical metasurface, using a constant boundary value to avoid problems with discontinuities at the boundary between meta-atoms. These gradient-based optimizations require the derivation of a shape derivative for the LSF boundary, also called the velocity field, which continuously reshapes the LSF towards an optimal solution for a given cost function.\cite{townsend_sensitivity_2013} The level set method has also been used to design dielectric metamaterials\cite{otomori_topology_2012} and metasurfaces\cite{dong2022} with the gradient-based adjoint variable method. Use of LSFs for \textit{gradient-free} optimization has been found in civil engineering with structural optimization of load-bearing structures such as bridges and cantilevers.\cite{guirguis_derivative-free_2016, guirguis_high-resolution_2018} 


In this paper, we present a method to improve metamaterial optimization that focuses on the design space parameterization. Our improved basis - the periodic level set function (P-LSF) shown in \autoref{fig:overview}(b) - boasts numerous benefits over the fragmented basis, including design space convexity, improved electromagnetic performance, and reduced iterations for convergence. The implicitly periodic nature of the P-LSF allows variation in material along the edges of the unit cell, in turn allowing freedom in optimizing continuity in currents and coupling of fields between unit cells. We use the P-LSF and fragmented parameterizations with finite-difference time-domain (FDTD) simulations to design two frequency selective surfaces: a bandpass filter and a high-Q notch filter. Gradient-free optimization is performed using evolutionary algorithms and parallelized with high-performance computing resources. We compare the two parameterizations based on optimization results for the two design objectives and finally manufacture a metasurface designed with the periodic level set method to measure its real-world performance.

\section*{Results}

\subsection*{Topology optimization for metasurfaces}

The metamaterial design problem can be formulated as a special case of a general topology optimization problem. 
We define a design region $\Omega\subset\R^D$ and a material distribution function $\rho\in L_2(\Omega)$ which is defined on $\Omega$. The vector space in which the distribution function resides is paired with a forward operator that maps distribution functions to points in another vector space which we call the "system response": $\mathcal{H} : L_2(\Omega)\rightarrow \R^M$, where $M$ is the number of samples.
We formulate a topology optimization problem to design a material distribution function minimizing the error between the resulting system response and a target response $\vy\in\R^M$ subject to inequality constraints $c_i:L_2(\Omega)\rightarrow\R$
\begin{equation}\label{eq:opt_problem}
\minimize_{\rho} \hspace{5pt} \sum_{m=1}^M \hspace{5pt}
L(\mathcal{H}_m(\rho), ~y_m)
\hspace{25pt}\textrm{subject to}\hspace{10pt}
c_i(\rho)\leq 0 \hspace{10pt}\forall i,
\end{equation}
where $L(\cdot, \cdot)$ is a loss function such as mean squared error. 

The general framework can be specialized to metasurfaces by taking $\Omega=\R^2$ and constructing a distribution function with a periodicity constraint of period $\vs=[s_1, s_2]$:
\begin{equation}\label{eq:periodicity}
\rho(\vx) = \rho(\vx + \begin{bmatrix}m*s_1\\n*s_2\end{bmatrix})
\hspace{20pt} \forall m,n\in\Z
\end{equation}
The region $\vx\in([0,s_1]\times[0,s_2])$ is representative of the infinite structure and is called the unit cell. The design of metasurfaces is challenging because the forward operator $\mathcal{H}$ often requires a computationally intensive full-wave electromagnetic simulation that does not compute gradients with respect to the material distribution function. To make metasurface optimization tractable, the material distribution function can be parameterized with a set of design variables that are tuned by an optimization algorithm.

\subsection*{Metasurface parameterization with level set functions}

\begin{figure}[ht]
\begin{center}
\includegraphics[width=\linewidth]{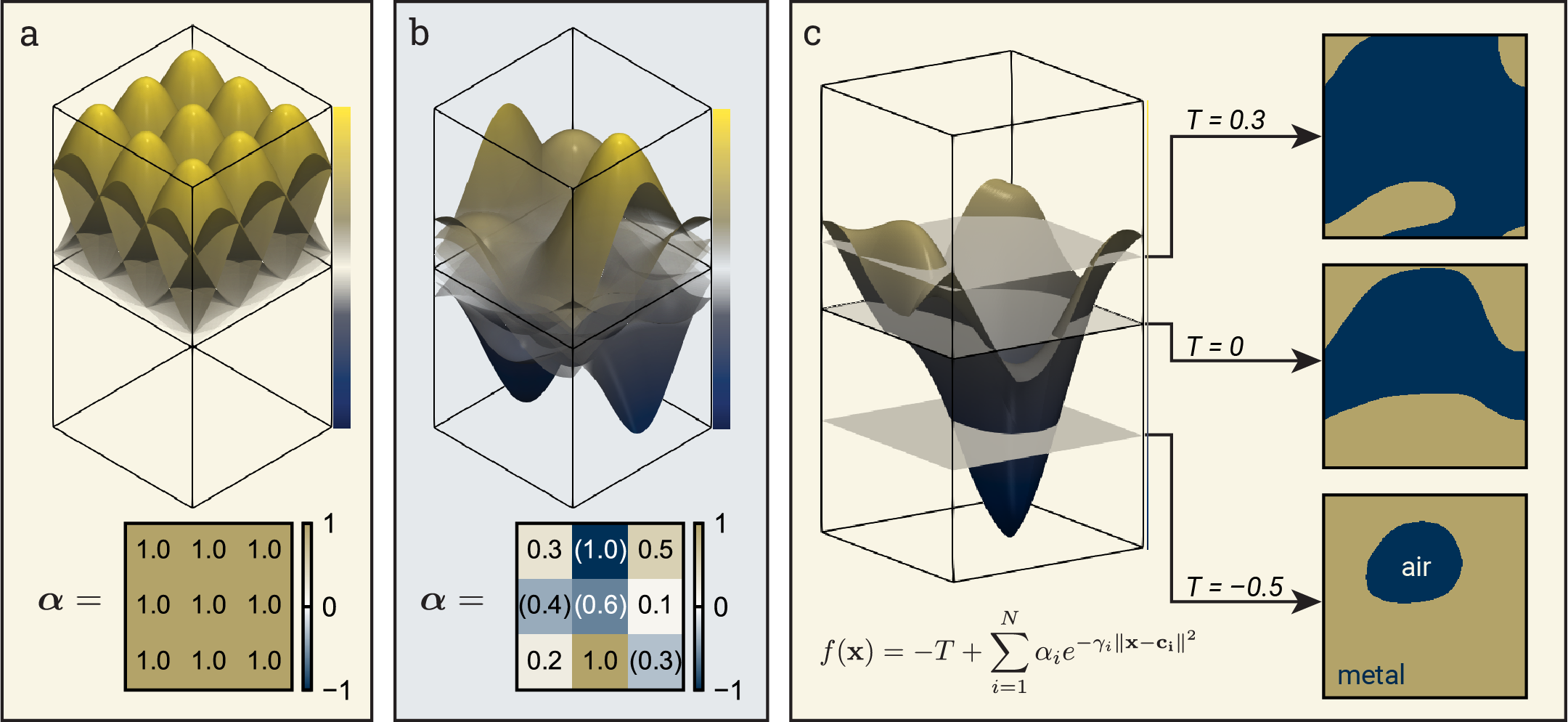}
\caption{The metasurface material distribution is determined by the level set of the sum of weighted radial basis functions (RBFs). (\textbf{a}) A set of $N$ RBFs is defined, with centers $\bm{c}_i$ such that they are spaced uniformly across a 2D plane.  (\textbf{b}) The set of radial basis functions is weighted by coefficients $\bm{\alpha}\in [-1, 1]$. (\textbf{c}) The sum of weighted RBFs plus the threshold term $T$ forms the level set function, and the material distribution on the 2D plane is determined by $f(\bm{x})>0$. Negative threshold values tend to produce material distributions dominated by metal, while those with large threshold levels have very little metal.}
\label{fig:test_rbf}
\end{center}
\end{figure}

The optimization of metamaterials and metasurfaces benefits from a design space parameterization which produces sophisticated material distribution functions while maintaining favorable properties for gradient-free optimization algorithms, i.e., a small number of design variables, a continuous and convex design space, and a Lipschitz continuous cost function. We address all of these goals simultaneously through level set functions.

For any scalar function $f(\vx)$ for $\vx\in\R^D$, a level set of $f$ is a hypersurface in $\R^D$ where the function takes on a constant value. For instance, given the 2-dimensional function $f(x, y)=\sqrt{x^2+y^2}$, the level set for any nonnegative level $f(x, y) = r$ is a circle of radius $r$. Without loss of generality, we will consider level set functions of the form $f(\vx) = g(\vx) - T$, where $T$ is the threshold, and the level set is always taken at $f(\vx)=0$. For the metasurface designs in this work, we define a binary material distribution of metal and air, where the zero level set defines the boundary between the two materials. However, the level set is applicable to continually varying material parameters, e.g. graded dielectrics\cite{giddens2020}, as well as the multiple material systems (air, resistive sheet, metal) found in absorbing metasurfaces\cite{feng2023}.




A common choice of level set function for topology optimization is composed of a sum of Gaussian radial basis functions (RBFs) with coefficients $\valpha$. Rounding is done relative to the zero level set.
\begin{equation}\label{eq:lsf}
f_{\valpha}(\vx) = -T+\sum_{i=1}^N \alpha_i e^{-\gamma_i^2 \|\vx-\vc_i\|^2} 
\hspace{40pt}
\rho_{\valpha}(\vx) = 
\begin{cases}
    1 \hspace{10pt}\textrm{ if } f_{\valpha}(\vx) > 0 \\
    0 \hspace{10pt}\textrm{ otherwise}
\end{cases}
\end{equation}
where $T$ is the threshold, $\vc_i$ are the RBF centers, and $\gamma_i$ are the scale factors (which are typically inversely proportional to the distance to adjacent centers). The norm used is nearly always the Euclidean $\mathcalboondox{l}_2$ norm, but we experiment with the $\mathcalboondox{l}_1$ and $\mathcalboondox{l}_{\infty}$ norms in this work as well. The basis coefficients and threshold determine the shape of the level set function and the resulting material distribution after rounding. For most problems, it is helpful to bound the basis coefficients since a single large positive or negative coefficient can result in a trivial one-material solution. \cite{guirguis_derivative-free_2016} \autoref{fig:test_rbf} shows examples for a uniformly spaced RBF basis, an LSF composed of a sum of RBFs, and the rounded material distribution functions for three different thresholds.




For topology optimization, the RBF basis coefficients $\valpha$ are the design variables presented to an optimization algorithm. The RBF parameters $\gamma_i$ and $\vc_i$ are initialized and kept fixed, and the threshold $T$ is often determined by a constraint on the volume of material in the material distribution function.\cite{guirguis_derivative-free_2016} For our metasurface design problems, we do not have a material constraint, so we simply fix the threshold at zero. For simulation and manufacturing, the parameterized material distribution function $\rho_{\valpha}$ is sampled at a set of grid points in $\vx$, which produces a binary image that we call the mask.

\subsection*{Periodic level set functions}

\begin{figure}
\begin{center}
\includegraphics[width=\linewidth]{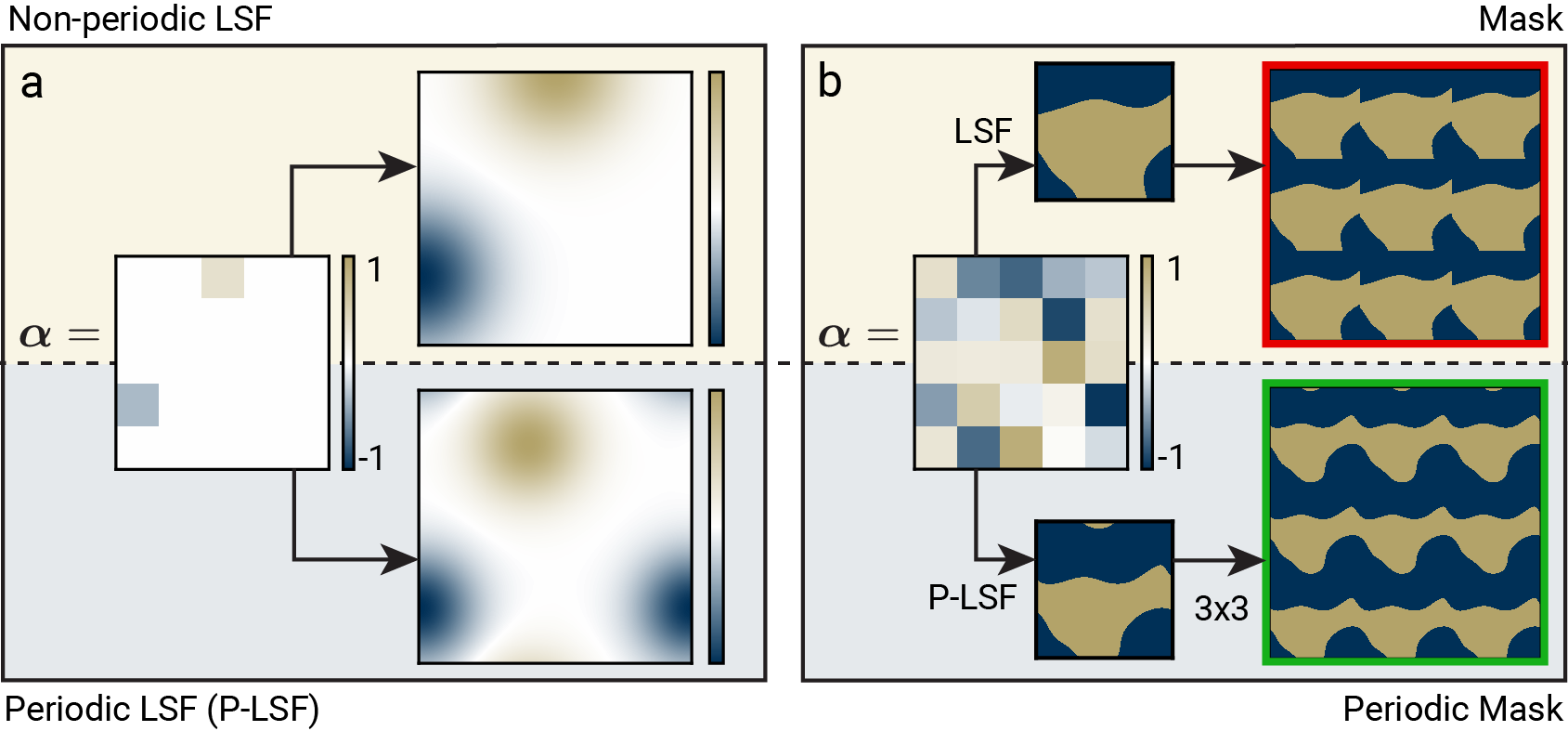}
\caption{ Level set functions (LSFs) used for periodic structures like metamaterials result in sharp material discontinuities on unit cell boundaries. We define the periodic LSF (P-LSF) which applies distance wrapping around unit cell boundaries to create continuous material features. (\textbf{a}) An LSF and P-LSF are defined with sparse basis coefficients, and the resulting function values are shown. (\textbf{b}) An LSF and P-LSF are defined with random basis coefficients and then rounded to binary-valued masks. When converted to a periodic structure, the LSF mask has material discontinuities on the unit cell boundaries, whereas the P-LSF mask removes the discontinuities. }
\label{fig:periodic_lsf_comparison}
\end{center}
\end{figure} 

To make level set functions more suitable for periodic structures like metamaterials, we propose a new formulation that wraps the distance from the topology variable $\vx$ to the RBF centers around the unit cell boundaries. Given an RBF center $\vc$ and a $D$-dimensional unit cell of size $\vs=[s_1,\hdots,s_D]$, we consider periodic shifts of $\vx$ to all other unit cells, and then take norm of the shift with the minimum distance to $\vc$. The wrapped $\mathcalboondox{l}_2$, $\mathcalboondox{l}_1$, and $\mathcalboondox{l}_{\infty}$ norms are defined as follows:




\begin{center}
\begin{tabular}{ | c | c | }

\hline
\textbf{Norm} & \textbf{Wrapped Norm} \\ \hline\hline

\parbox[c][10ex][c]{1.5cm}{\centering$\|\vx-\vc\|_2$} & 
\parbox[c][10ex][c]{6cm}{\centering\[\sqrt{\sum_{d=1}^D \hspace{5pt} \min_{n \in \mathbb{Z}} \hspace{5pt} (x_d + n * s_d - c_d)^2}\]}
\\ \hline

\parbox[c][10ex][c]{1.5cm}{\centering$\|\vx-\vc\|_1$} & 
\parbox[c][10ex][c]{6cm}{\centering\[\sum_{d=1}^D \hspace{5pt} \min_{n \in \mathbb{Z}} \hspace{5pt} |x_d + n * s_d - c_d|\]}
\\ \hline

\parbox[c][10ex][c]{1.5cm}{\centering$\|\vx-\vc\|_{\infty}$} & 
\parbox[c][10ex][c]{6cm}{\centering\[\max_{d=1}^D \hspace{5pt} \min_{n \in \mathbb{Z}} \hspace{5pt} |x_d + n * s_d - c_d|\]}
\\ \hline
\end{tabular}
\end{center}

The proposed distance wrapping enables RBF features to flow continuously around the boundaries of the unit cell. Without this modification, material features do not align across the unit cell boundaries, which results in discontinuities when the LSF is converted to a periodic structure as shown in \autoref{fig:periodic_lsf_comparison}(b). In addition to distance wrapping, the spacing of RBF centers must account for wrapping as well to avoid crowding RBFs at the edges of the unit cell. We term the LSF formulation with distance wrapping the ``periodic level set function'' (P-LSF).

We note that sinusoids could also be used to construct level set functions with inherent periodicity on the unit cell. Such a basis would cause each parameter to impact material distribution globally across the unit cell with larger changes from lower-frequency basis functions; use of RBFs with added periodicity maintains the locality of each parameter. For small changes in basis coefficients, a cost function varies less (i.e., has a smaller Lipschitz constant) due to localized material placement with RBFs rather than global material placement with sinusoids.

\subsection*{Upsampling level set function coefficients}

An effective approach to gradient-free topology optimization is multi-stage optimization, where an initial coarse problem with a reduced design space is solved quickly and iteratively refined to a fine-grained solution.\cite{guirguis_derivative-free_2016, guirguis_high-resolution_2018} We experiment with multi-stage optimization for metasurfaces with the P-LSF basis, which requires upsampling coefficients from coarse to fine bases. We construct a coarse P-LSF basis as a set of $M$ RBF functions $\{\phi_m\}$ centered at $\{\vc_m\}$ with basis coefficients $\valpha\in\R^M$. Likewise, we construct a fine basis with $N$ RBF functions $\{\psi_n\}$ centered at $\{\vc'_n\}$ with basis coefficients $\vbeta\in\R^N$. The goal of upsampling is to take a fixed coefficient vector $\valpha$ from the coarse basis and find a suitable coefficient vector $\vbeta$ in the fine basis which most closely emulates the level set function from the coarse basis coefficients.




We first attempt the matrix inverse method proposed by Guirguis et al. \cite{guirguis_high-resolution_2018} to upsample P-LSF basis coefficients. The method attempts to match the coarse and fine P-LSF values at the $N$ centers for the fine basis, which results in a system of $N$ equations to solve for the $N$ fine coefficients.
We first construct a vector $\vy$ of length $N$ by sampling the coarse P-LSF at the center location of each basis function for the fine P-LSF:
\begin{equation}\label{eq:upsample_y}
\vy =
\begin{bmatrix}
\sum_m \alpha_m\phi_m(\vc'_1), & \sum_m \alpha_m\phi_m(\vc'_2), & \hdots, & \sum_m \alpha_m\phi_m(\vc'_N)
\end{bmatrix}^T
\end{equation}
We also construct an $N\times N$ sampling matrix $\mG$ for the fine basis, and  vector $\vbeta$ of the unknown fine basis coefficients:
\begin{equation}\label{eq:upsample_inv_matrix}
\mG = 
\begin{bmatrix}
    \psi_1(\vc'_1) & \psi_2(\vc'_1) & \hdots & \psi_N(\vc'_1) \\
    \psi_1(\vc'_2) & \psi_2(\vc'_2) & \hdots & \psi_N(\vc'_2) \\
    \vdots & \vdots & \ddots & \vdots \\
    \psi_1(\vc'_N) & \psi_2(\vc'_N) & \hdots & \psi_N(\vc'_N) \\
\end{bmatrix}
\hspace{20pt}
\vbeta =
\begin{bmatrix}
\beta_1, & \beta_2, & \hdots, & \beta_N
\end{bmatrix}^T
\end{equation}
We intend to solve the system of equations $\mG\vbeta=\vy$, so we find the fine basis coefficients by inverting $\mG$ and taking $\vbeta = \mG^{-1}\vy$. 

The matrix inverse method works well when the $\mathcalboondox{l}_2$ norm is used for the radial basis functions, but it fails when applied to upsampling with the $\mathcalboondox{l}_1$ and $\mathcalboondox{l}_{\infty}$ norms, where the resulting sampling matrix is ill-conditioned and results in large upsampled basis coefficients. We instead derive a more robust P-LSF upsampling process using an overdetermined system of equations and the pseudo-inverse of the resulting rectangular sampling matrix.

\begin{figure}[ht]
\begin{center}
\includegraphics[width=\textwidth]{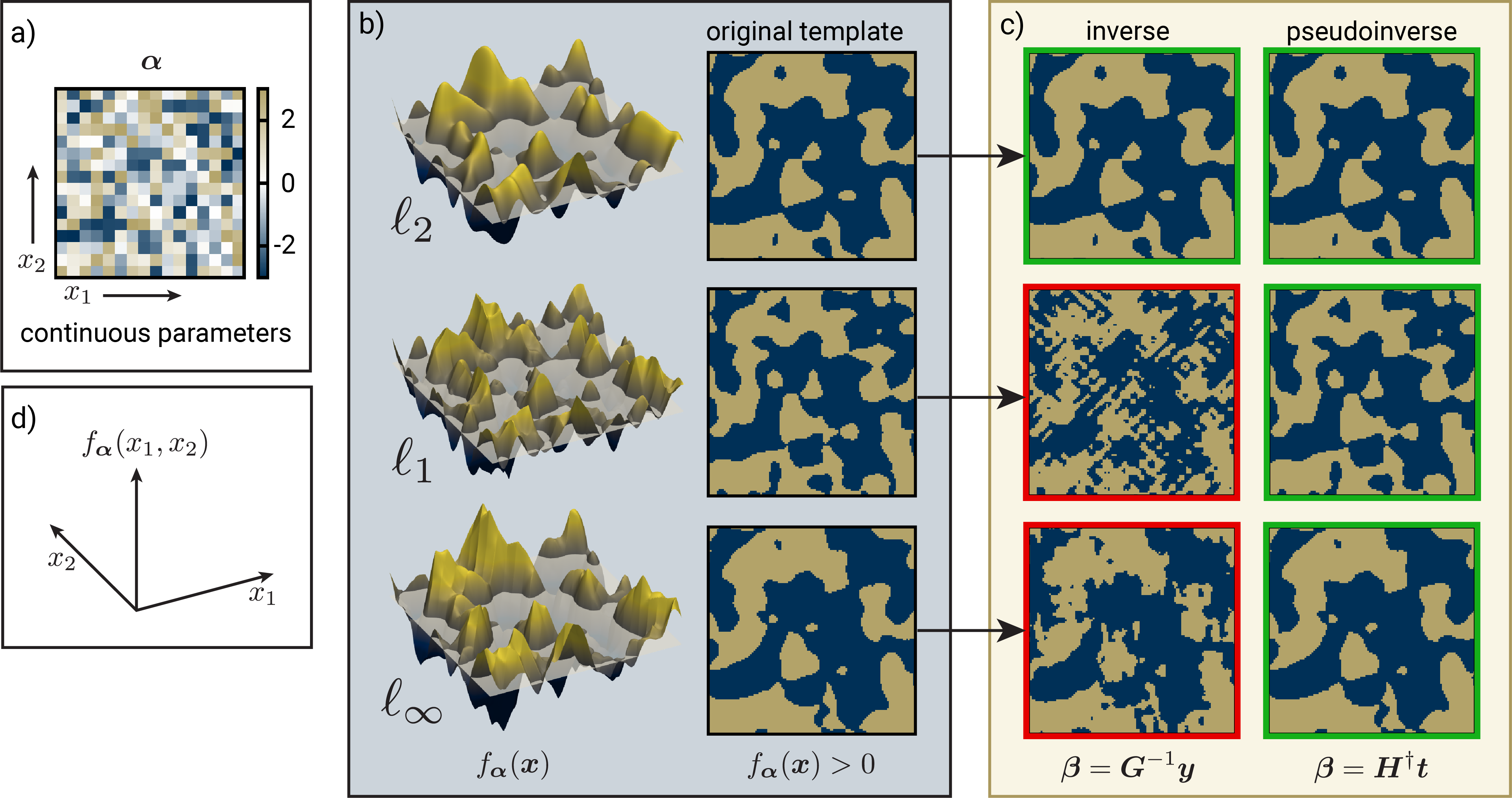}
\caption{A $16\times16$ P-LSF is upsampled to $32\times32$. \textbf{a}) A set of 256 parameters in $[-1,1]$ is randomly chosen. \textbf{b}) P-LSF functions are created with the $\mathcalboondox{l}_2$, $\mathcalboondox{l}_1$, and $\mathcalboondox{l}_{\infty}$ norms. The P-LSFs are binarized based on the zero level. \textbf{c}) The proposed pseudo-inverse upsampling method is artifact-free, while the matrix inverse method\cite{guirguis_high-resolution_2018} is not, especially in the $\mathcalboondox{l}_1$ and $\mathcalboondox{l}_{\infty}$ norms. \textbf{d}) Coordinate system.}
\label{fig:upsample_examples}
\end{center}
\end{figure}

The pseudo-inverse method solves a least-squares problem matching the coarse and fine P-LSF values at $P>N$ mask positions $\{\vz_p\}$, instead of at the $N$ fine basis centers. The number of mask positions to sample may be arbitrarily selected; the spacing in a uniformly spaced grid of mask positions is generally chosen based on the resolution required for a discretized simulation or manufacturing process. We first construct a vector $\vt$ by sampling the coarse basis at the $P$ mask positions $\vz_p$, which is equivalent to the coarse P-LSF mask values before rounding:
\begin{equation}\label{eq:upsample_t}
\vt =
\begin{bmatrix}
\sum_m \alpha_m\phi_m(\vz_1), & \sum_m \alpha_m\phi_m(\vz_2), & \hdots, & \sum_m \alpha_m\phi_m(\vz_P)
\end{bmatrix}^T
\end{equation}
We then construct a $P\times N$ sampling matrix $\mH$ and solve for the upsampled coefficients $\vbeta$ with the pseudo-inverse:
\begin{equation}\label{eq:upsample_pinv_matrix}
\mH = 
\begin{bmatrix}
    \psi_1(\vz_1) & \psi_2(\vz_1) & \hdots & \psi_N(\vz_1) \\
    \psi_1(\vz_2) & \psi_2(\vz_2) & \hdots & \psi_N(\vz_2) \\
    \vdots & \vdots & \ddots & \vdots \\
    \psi_1(\vz_P) & \psi_2(\vz_P) & \hdots & \psi_N(\vz_P) \\
\end{bmatrix}
\hspace{30pt}
\vbeta = \mH^{\dagger}\vt
\end{equation}

We find that this formulation works well for both the $\mathcalboondox{l}_1$ and $\mathcalboondox{l}_{\infty}$ LSF upsampling. Additionally, we find the pseudo-inverse method matches masks with the $\mathcalboondox{l}_2$ norm more closely than the matrix inverse method. Upsampling examples for masks with each norm are shown in \autoref{fig:upsample_examples}.




\subsection*{System and cost function construction}

As a case study, we construct representative geometries for a microwave metasurface using both fragmented and periodic level set function parameterizations and optimize them for two different objectives: a broadband 2:1 bandpass frequency-selective filter and a multi-resonance narrowband one. Optimization is performed solely on the material distribution function for a thin metal layer on dielectric substrate. For fragmented designs, we divide a $120\times120$ metal mask into staggered fragments of size $4\times4$ pixels which results in $N=915$ binary design variables. To complement this, we define a P-LSF basis composed of $32\times32$ RBF functions arranged uniformly across the unit cell for a total of $N=1024$ continuous design variables bounded in the range $[-1,1]$. The metasurface optimization problem is:
\begin{equation}\label{eq:metasurface_opt}
\maximize_{\valpha} \hspace{5pt} \sum_{m=1}^M \hspace{5pt}
L(\mathcal{H}_m(\rho_{\valpha}), ~y_m)
\hspace{15pt}\textrm{subject to}
\begin{cases}
-1\leq\alpha_i\leq1, \hspace{10pt} & 1\leq i\leq N\hspace{10pt}\textrm{(P-LSF)} \\
\alpha_i\in\{0,~1\},               & 1\leq i\leq N\hspace{10pt}\textrm{(fragmented)}
\end{cases}
\end{equation}

We use the maximization convention to align with genetic algorithm notion of maximizing a fitness function. We construct the forward operator $\mathcal{H}$ by first mapping the parameterized material density function $\rho_{\valpha}$ to a metal mask on the simulated metasurface geometry. We perform a full-wave electromagnetic simulation of an incident plane wave followed by a near-field to far-field transform and Fourier transform to compute the metasurface scattering parameters (S-parameters) at $M$ different frequencies. The resulting S-parameters are scored with a target system response $\vy$ and loss function $L$ designed to describe the desired performance of the metasurface for each objective.


\begin{figure}
\begin{center}
\includegraphics[width=\linewidth]{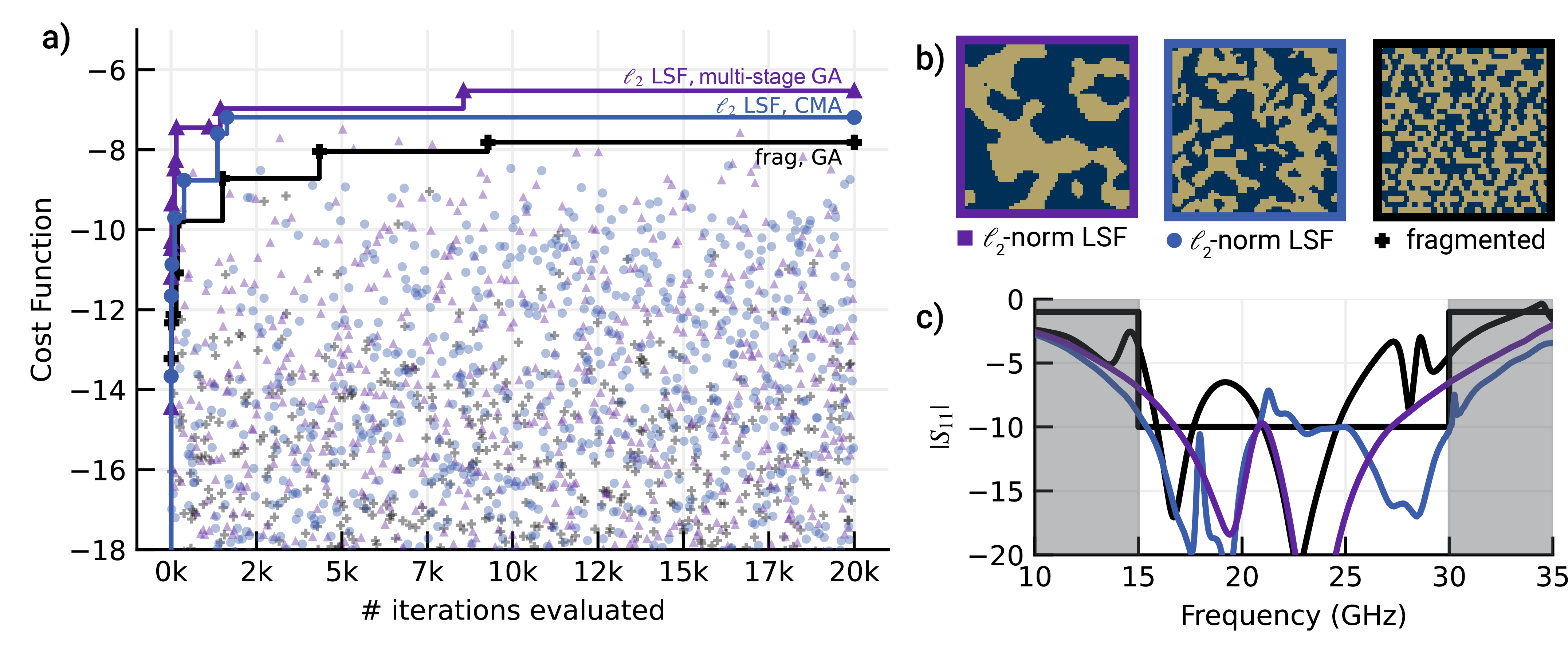}
\caption{\textbf{a}) For the broadband design objective, the LSF method converged more quickly and to a slightly higher value than the fragmented method. Lines indicate best function evaluations over the course of the optimization, while the scatter plot shows a decimated sampling of the scores for all function evaluations. \textbf{b}) The resulting best mask from each optimization is displayed. \textbf{c}) Reflection coefficient for each best case.}
\label{fig:opt_convergence_wideband}
\end{center}
\end{figure}

The first objective aims to design a 2:1 bandpass filter which minimizes the reflectance ($S_{11}$) between the passband frequency range of 15-30~GHz while maximizing the reflectance outside this band. We use the weighted mean absolute error (MAE) as the loss function and define the target system response as an $S_{11}$ of less than -10~dB in the passband and greater than -0.5~dB in the stopband. $S_{11}$ values exceeding the design goal are considered to have an error of zero and hence optimal. We negate this objective to construct the maximization problem.

As a second objective, we design a highly resonant metasurface for materials characterization.\cite{howard2022loss} We specify two frequency bands from 8-16~GHz and 24-32~GHz and maximize the Q-factor of one notch filter in each band. Techniques for Q-factor computation using half-power bandwidth or Q-circle fitting\cite{gregory2022} can be unstable for frequency responses with closely spaced resonances, so we compute a cost function which emulates Q-factor maximization. We construct a design-dependent loss function which selects the frequency of minimum transmittance ($S_{21}$) in each design band as the predicted notch location. The cost function value is computed as the MAE between the minimum and average transmittance in a 1~GHz range around the two predicted notch locations, which is to be maximized. $S_{21}$ values outside of the range are ignored. We find that this cost function encourages the optimizer to both maximize notch depth and minimize the notch width, which effectively maximizes the Q-factor.

\begin{figure}[hbt]
\begin{center}
\includegraphics[width=5.63in]{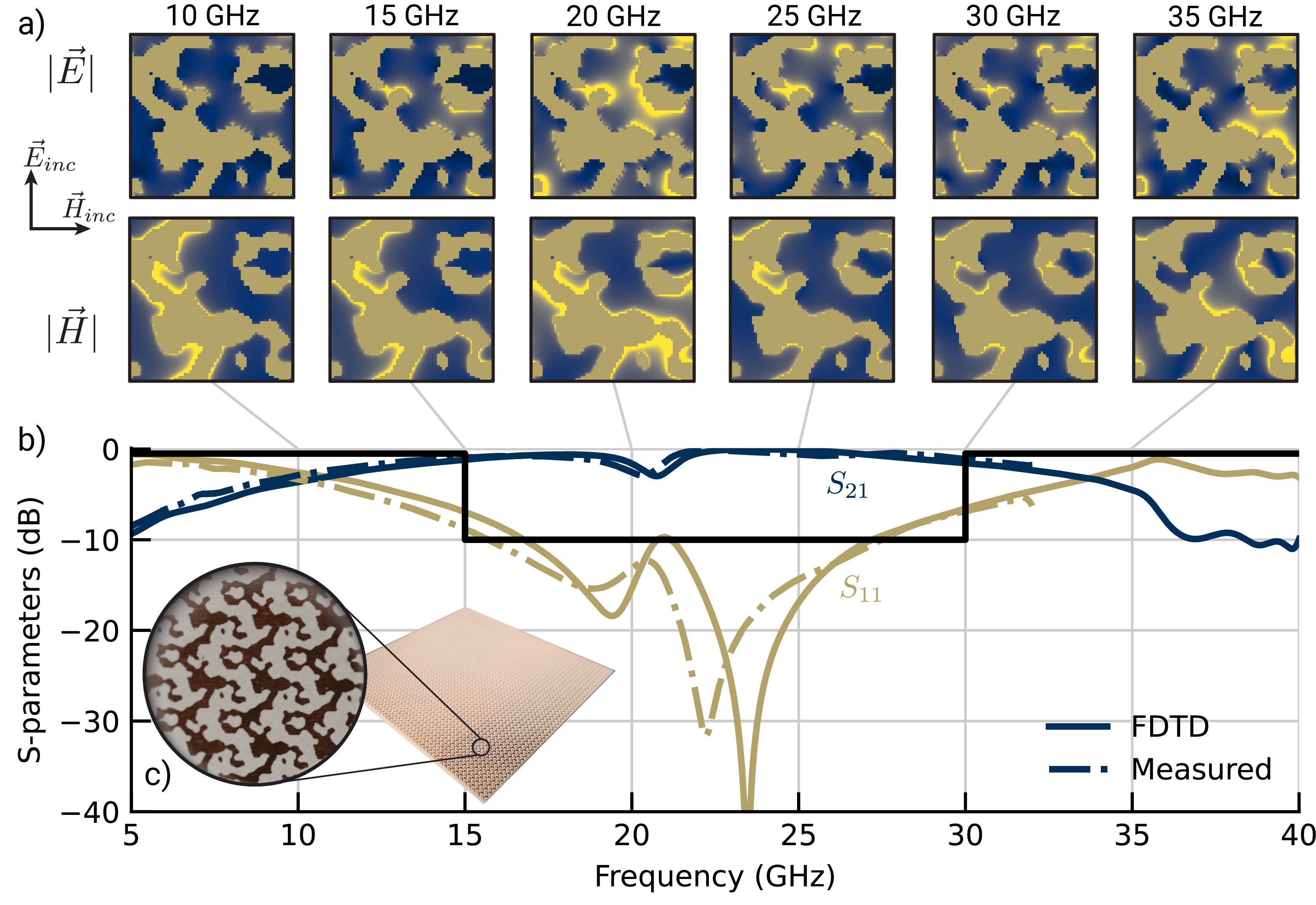}
\caption{Broadband 2:1 bandpass metasurface. \textbf{a}) Field maps generated from FDTD data are shown at discrete frequencies. \textbf{b}) S-parameters measured in the focused beam system show good agreement with simulated FDTD outputs. \textbf{c}) The fabricated metasurface is shown with a detailed view of a few periods of the structure.}
\label{fig:bp_lsf_sparameters}
\end{center}
\end{figure}

\subsection*{Optimization algorithms}

We apply the non-dominated sorting genetic algorithm II\cite{deb2002} (NSGA-II) to optimize microwave metasurface designs with both P-LSF and fragmented parameterizations for both bandpass and notch objectives. To improve algorithm convergence, we also implement multi-stage optimization with both design parameterizations, which divides optimization into stages with incrementally higher granularity. In this work, we apply a two-stage optimization procedure. The first stage of optimization uses a reduced number of coarse-grained design variables to quickly converge to an acceptable solution set. The coarse solutions are upsampled to a finer basis and used as seeds for a fine-grained optimization to find the final solution. For both fragmented and P-LSF parameterizations, the coarse basis has approximately $25\%$ the number of parameters as the fine basis. Though the coarse optimization stage may bias the fine stage towards a locally optimal solution, this is favorable since optimizing directly in the fine basis may not converge in the allocated time.

The continuous design space of the P-LSF parameterization lends itself to a variety of optimizers other than genetic algorithms. We use the package Nevergrad \cite{nevergrad} to flexibly test various single- and multi-threaded optimization algorithms and found the most success with the covariance matrix adaptation evolution strategy (CMA-ES). \cite{hansen_reducing_2003} CMA-ES is an effective gradient-free optimizer for nonlinear functions on continuous domains which models the acquisition function with a multivariate Gaussian distribution. As progressive function evaluations are acquired, the Gaussian covariance matrix is updated and used to select successive sample points. CMA-ES has previously been applied to optimizing geometric parameters for antenna design. \cite{gregory_fast_2011}

A comparison between the convergence of different optimization techniques for the bandpass design objective is shown in \autoref{fig:opt_convergence_wideband}. The metasurface designed with the P-LSF basis and two-stage optimization achieves the best cost function value, followed closely by the P-LSF metasurface with one-stage optimization. Though applying two-stage optimization did not significantly increase the final bandpass cost function value, two-stage optimization converged closely to the final value in approximately 1/3rd the iterations as one-stage optimization. Applying the CMA-ES optimizer to the P-LSF metasurface did not fully converge within the allocated function evaluations but still produced a comparable P-LSF design. The fragmented metasurfaces with both one-stage and two-stage optimization perform worse than the P-LSF based metasurfaces for the bandpass design objective. For the fragmented basis, introducing two-stage optimization showed improvement in the final objective value as well as accelerate convergence to the final value.

\begin{figure}
\includegraphics[width=\linewidth]{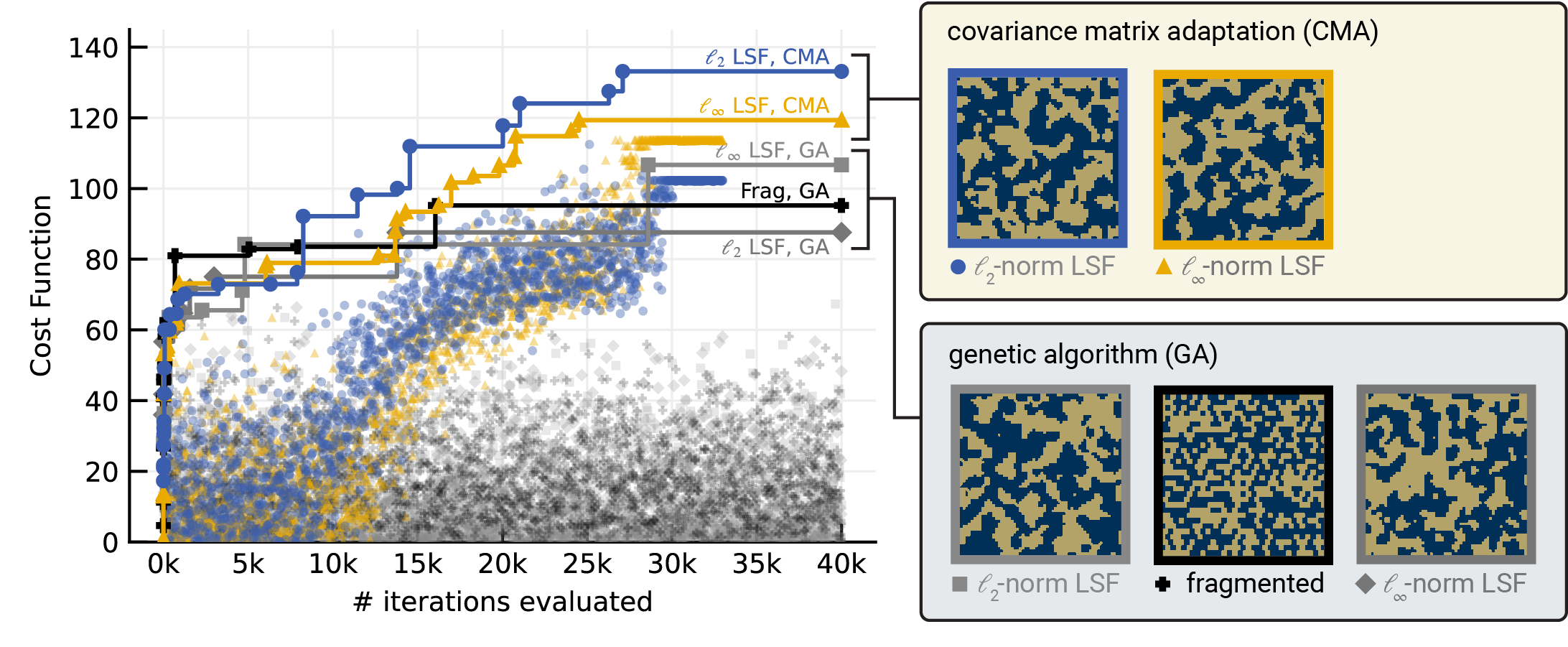}
\caption{The convergence plot for the design of a dual-band high-Q notch filter shows improved performance using the LSF method in conjunction with the CMA-ES optimizer. Lines indicate best function evaluations over the course of the optimization, while the scatter plot shows a decimated sampling of the scores for all function evaluations.}
\label{fig:opt_convergence_narrowband}
\end{figure}

Convergence for the notch objective is shown in \autoref{fig:opt_convergence_narrowband}. Applying two-stage optimization to the P-LSF based metasurface offered little benefit and slowed down optimization convergence. Similarly for the fragmented designs, two-stage optimization detracted from the final achieved cost function value, likely from premature convergence to a local optimum. The fragmented design outperformed the original P-LSF design, so we attempted a different P-LSF parameterization. As opposed to using the $\mathcalboondox{l}_2$ norm in the RBFs constituting the P-LSF which results in round material features, we instead apply the $\mathcalboondox{l}_{\infty}$ norm, which tends towards rectangular features similar to fragmented designs. Optimizing the $\mathcalboondox{l}_{\infty}$ P-LSF metasurface resulted in a design that performed better than the best fragmented design. Additionally, we applied the CMA-ES optimizer to both the $\mathcalboondox{l}_{2}$ and $\mathcalboondox{l}_{\infty}$ norm P-LSF metasurfaces. CMA-ES improved the final cost function value in both cases, and the resulting $\mathcalboondox{l}_{2}$ norm design substantially outperformed the $\mathcalboondox{l}_{\infty}$ norm design.

\begin{figure}[hbt]
\includegraphics[width=5.63in]{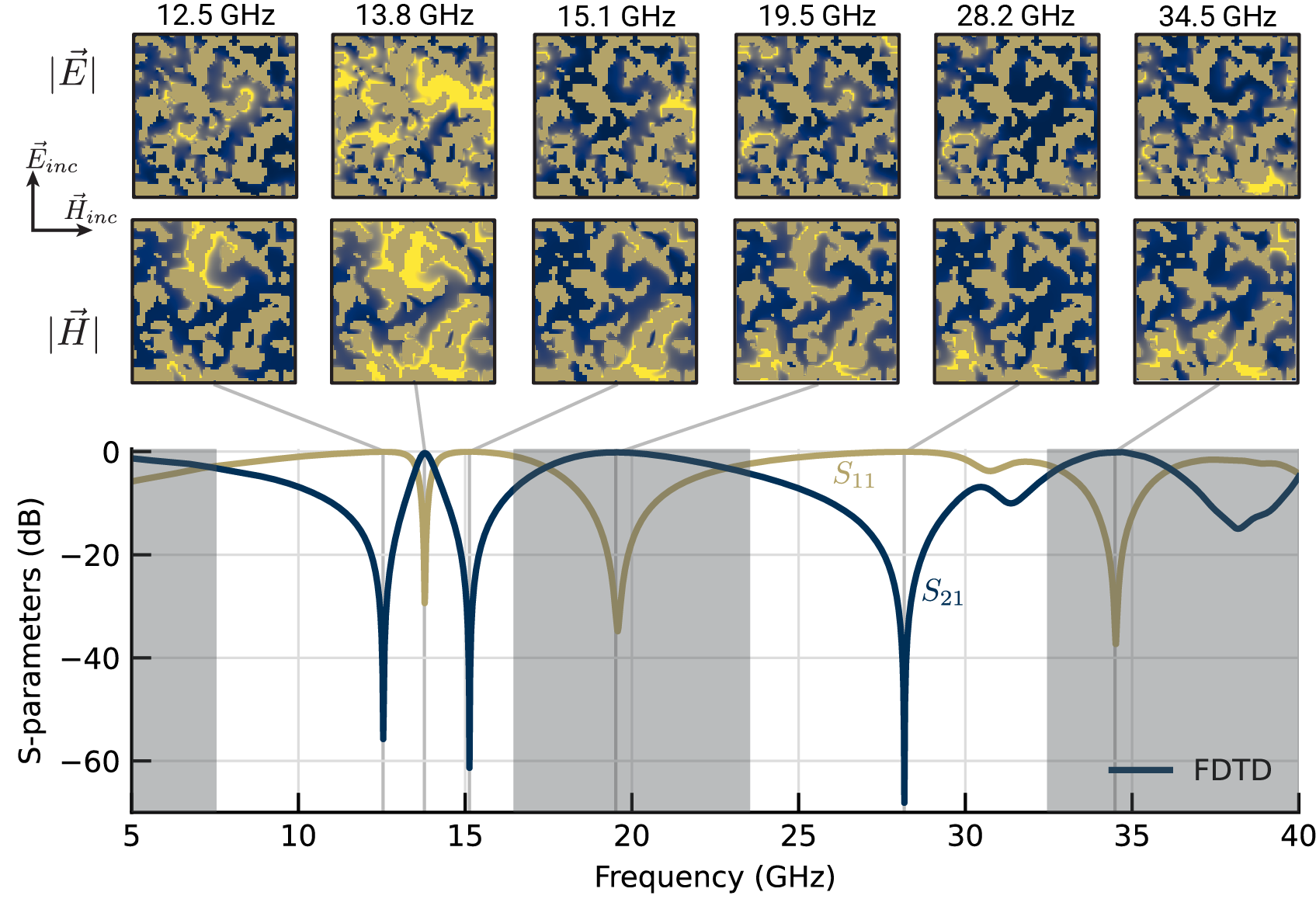}
\caption{Optimization convergence and S-parameters for the dual-band high-Q notch filter. The cost function is the difference between the minimum and average transmittance ($S_{21}$) in a 1~GHz band around one notch in each frequency band.}
\label{fig:narrowband_lsf_sparameters}
\end{figure}

\subsection*{Measurement}

To validate the design process, we fabricated the optimized 2:1 bandpass filter metasurface suggested by the P-LSF GA multi-stage optimization. We measured transmission and reflection scattering parameters in a free-space focused beam system, and close agreement between the full-wave simulation and measured results is shown in \autoref{fig:bp_lsf_sparameters}.

\section*{Discussion}

For topology optimization problems, the interplay between material distribution functions and the design variables presented to gradient-free optimization algorithms determines algorithm convergence and the performance of optimized designs. Canonical metasurface designs - Jerusalem cross, split-ring resonator, etc.\cite{islam2022, kent2010new, monavar2011bandwidth} - offer well-behaved, convex design spaces but are too heavily constrained to push the boundaries of achievable frequency responses. The fragmented parameterization offers many more degrees of freedom, but has a discrete, non-convex design space (the vertices of a unit hypercube, $\valpha\in\{0, 1\}^N$) which is completely disjoint and poses a difficult combinatorial optimization problem. A new basis for periodic structures like metasurfaces is defined using the periodic level set function (P-LSF), which creates a continuous, convex design space (the interior of a hypercube, $\valpha\in[-1, 1]^N$) that facilitates optimization. Moving from a non-convex to a convex design space improves the performance of applied optimization algorithms. \cite{rockafellar_lagrange_1993, boyd_convex_2004} Genetic algorithms in particular benefit from a continuous, convex design space since the mutation and mating operators can perturb and interpolate between individuals more effectively,\cite{haupt_practical_2004} and our results demonstrate accelerated genetic algorithm convergence with the P-LSF basis.

The P-LSF basis also produces more sophisticated material distribution functions than either the fragmented basis or canonical designs. Considering bases with the same number of parameters, a fragmented basis can only construct rectangular material features which are integer multiples of the fragment size, whereas a P-LSF basis can realize continuously varying features as well as diagonal or rounded edges that are unattainable in the fragmented basis. Masks constructed with the P-LSF basis can be arbitrarily resampled at higher or lower resolutions to account for simulation regridding or manufacturing constraints on the mask. As shown in \autoref{fig:opt_convergence_wideband} and \autoref{fig:opt_convergence_narrowband}, the P-LSF basis can realize both large and small material features as guided by the design objective, whereas the fragmented basis is restricted in the range of achievable features.

Using the $\mathcalboondox{l}_2$ norm when defining the P-LSF results in rounded features which are suitable for broadband cost functions. For narrowband problems, the P-LSF basis can produce sharper features by redefining the basis with the $\mathcalboondox{l}_{\infty}$ norm. Multi-stage optimization can be applied to both P-LSF and fragmented designs to accelerate convergence by solving an initial coarse problem and incrementally refining the coarse solution to a fine one.

The P-LSF can be used as a drop-in replacement for fragmented designs with few changes to existing evolutionary algorithm infrastructure. Additionally, the P-LSF facilitates the use of more efficient optimization algorithms such as the covariance matrix adaptation evolution strategy (CMA-ES) which can converge to better solutions in fewer iterations. The P-LSF method is broadly applicable to array technologies including microwave spatial filters as shown in this work, optical elements for super-resolution imaging\cite{shi2023}, graded metasurfaces,\cite{huang2023} and antenna arrays.\cite{mair2023}

\section*{Methods}

\subsection*{FDTD Modeling} 

We performed electromagnetics simulations using FDTD software developed at Georgia Tech Research Institute (GTRI). We discretized each metasurface design onto a $\bm{N} = [120, 120, 62]^T$ grid with a uniform grid cell size of \SI{0.051}{\mm} (\US{0.002}{\inch}) in each dimension. The resulting periodicity of the proposed metasurface is \SI{6.1}{\mm} (\US{0.240}{\inch}) in both planar dimensions. The supporting substrate for the metasurface is chosen to be \SI{.127}{\mm} (\US{0.005}{\inch}) Rogers RO3003 ($\epsilon_r'=3.0\pm0.04, \tan{\delta}=0.001$)\cite{rogers3003}. Periodic boundary conditions were used on $x-$ and $y-$faces, and the simulation excited with a normally incident differentiated Gaussian pulse ($f_p = \SI{15}{\GHz}$) plane wave. We ran each simulation for $8,000$ time steps ($\Delta t = \SI{95.876}{\fs}$) and used Prony's method to extrapolate accurate frequency domain data from any residual energy due to resonance in the grid.\cite{ko1991} The imaginary permittivity of the dielectric substrate was modeled in FDTD as a constant conductivity $\sigma = \SI{0.00167}{\siemens\per\meter}$.


\subsection*{Optimization parameters}

For the genetic algorithm optimizations, we define a mating operator for both the fragmented and P-LSF parameterizations with uniform crossover for binary parameters and intermediate recombination ($\alpha=0.5$) for continuous parameters. For fragmented designs, we define an entry-wise binary mutation with a $10\%$ probability of flipping the value of each fragment. For the P-LSF, we define an entry-wise continuous mutation which has a $10\%$ chance of triggering for each individual RBF basis coefficient. If a mutation is triggered, there is an $80\%$ chance to apply a uniform random mutation which is bounded in a small range around the current basis coefficient, and a $20\%$ chance to redraw the basis coefficient from the full range of values. We use a population size of 200 and a total of 200 generations, or $40,000$ total cost function evaluations.

CMA-ES is applied to the P-LSF designs using the \texttt{ParameterizedCMA} implementation found in the Nevergrad package\cite{nevergrad}, which wraps the \texttt{pycma} algorithm.\cite{hansen_cma-espycma_2023}. The default optimization parameters were used. A total of $40,000$ cost function evaluations were performed, but we found no improvement after $10,000$ evaluations. 

\subsection*{Fabrication and S-parameter measurement}

The broadband metasurface was made by printed circuit board processes by etching the pattern on \SI{.127}{\mm} (\US{0.005}{\inch}) thick Rogers RO3003 ($\epsilon_r'=3.0\pm0.04, \tan{\delta}=0.001$).\cite{rogers3003} with \textonehalf~oz. electrodeposited copper foil. The final finish of the board was bare copper. 

Scattering parameters for the metasurface were measured in GTRI's free-space focused beam system as described by Howard et al.\cite{howard2022loss} The time gate width used for the Gated Reflect Line\cite{bartley2005} calibration was \SI{4}{\ns}.

\nolinenumbers

\bibliography{references}

\section*{Acknowledgements}


The authors thank Sam McWhorter for his help in layout for fabrication of the metasurface and Amanda Kieffer for assistance with graphics. This work was supported by Georgia Tech Research Institute's Independent Research and Development program, for which the authors are grateful. 

\section*{Author contributions statement}

A.S. developed the parameterization approach and performed optimizations. C.H. designed FDTD models, and measured the fabricated metasurface. A.S. and C.H. wrote the manuscript, with C.H. designing the figures. K.A. and J.R. advised on the research and analyzed results. All authors reviewed the manuscript.

\section*{Additional information}

\textbf{Competing interests:} The authors declare no competing interests.

\end{document}